\documentclass[namedreferences]{SolarPhysics}
\usepackage{epsfig}

\newcommand{\be}{\begin{equation}}
\newcommand{\ee}{\end{equation}}
\newcommand{\beq}{\begin{eqnarray}}
\newcommand{\eeq}{\end{eqnarray}}

\begin{document}
\begin{article}
\begin{opening}
\input{psfig.tex}
\title{The Ly$\alpha$ and Ly$\beta$ Profiles in Solar Prominences and Prominence Fine Structure }
\author{J.-C. Vial$^{1}$, H. Ebadi$^{1,2,3}$, A. Ajabshirizadeh$^{2,3}$ \\}
\institute{$^{1}$ Institut d'Astrophysique Spatiale, Unit\'e Mixte\\
CNRS-Universit\'e de Paris XI, Bat 121, 91405 Orsay, France\\
$^{2}$Tabriz University, Faculty of Physics, Tabriz, Iran\\
$^{3}$Research Institute for Astronomy and Astrophysics of Maragha,\\
55134-441 Maragha, Iran\\}
\begin{abstract}
Ly$\alpha$ and Ly$\beta$ line profiles in a solar prominence were
observed with high spatial and spectral resolution with SOHO/SUMER.
Within a 60 arcsec scan, we measure a very large variety of
profiles: not only reversed and non-reversed profiles but also
red-peaked and blue-peaked ones in both lines. Such a spatial
variability is probably related to both the fine structure in
prominences and the different orientations of mass motions.
 The usage of integrated-intensity cuts along the SUMER slit, allowed us to
categorize the prominence in three regions.
 We computed average profiles and integrated intensities in these lines which are
 in the range (2.36 -- 42.3) W m$^{-2}$ sr$^{-1}$ for Ly$\alpha$
 and (0.027 -- 0.237) W m$^{-2}$ sr$^{-1}$
for Ly$\beta$. As shown by theoretical modeling, the
Ly$\alpha$/Ly$\beta$ ratio is very sensitive to geometrical and
thermodynamic properties of fine structure in prominences. For some
pixels, and in both lines, we found agreement between observed
intensities and those predicted by one-dimensional models. But a
close examination of the profiles indicated a rather systematic
disagreement concerning their detailed shapes. The disagreement
between observations and thread models (with ambipolar diffusion)
leads us to speculate about the importance of the temperature
gradient between the cool and coronal regions. This gradient could
depend on the orientation of field lines as proposed by Heinzel,
Anzer, and Gun\'{a}r (2005).
\end{abstract}

\end{opening}
\section{Introduction.}
Hydrogen lines are the most prominent lines observed in solar
prominences. The resonance lines of the Lyman series have been
observed since the Skylab ATM experiment. The full profiles of the
Ly$\alpha$ and Ly$\beta$ lines were obtained with the UV
polychromator on OSO 8 for the first time in a quiescent prominence
by Vial, 1982a.
 Both were reversed and the Ly$\alpha$ intensity was about equal to the
incident chromospheric intensity multiplied by the dilution factor.
The opacities are very high and radiation transfer is dominated by
the scattering of chromospheric Ly$\alpha$ and Ly$\beta$ photons.
The Ly$\alpha$ line has been extensively used as a diagnostic tool
in the quiet or active chromosphere and especially in solar
prominences (Vial, 1982b). Schmieder \emph{et al}. (1999) and
Heinzel \emph{et al.} (2001) presented a nearly simultaneous
observation of the whole Lyman series including the Ly$\alpha$ and
Ly$\beta$ lines. But the Ly$\alpha$ line profile was affected by the
detector attenuator.
\\The Ly$\alpha$/Ly$\beta$ ratio is very sensitive to the physical and geometrical
properties of fine structures and consequently it provides a
diagnostic tool for deriving the fine structure of solar prominences
(Vial \emph{et al}., 1989; Rovira \emph{et al}., 1994; Fontenla
\emph{et al}., 1996; Heinzel \emph{et al}., 2001). Non-LTE radiative
transfer modeling of prominences using plane-parallel infinite slabs
by Gouttebroze, Heinzel, and Vial (1993, hereafter GHV) yielded
large values for this ratio (90 to 400) contrary to the OSO 8
observed value (65). Fontenla and Rovira (1983, 1985) and Vial
\emph{et al}. (1989) constructed thread models and solved
simultaneously the radiative transfer, statistical equilibrium, and
ionization equations assuming a three-level atom plus continuum.
Their results showed that the Ly$\alpha$ intensities are in
agreement with observations, but the Ly$\beta$ line intensities are
too small compared with those observed by OSO 8 (Vial, 1982a). For
strongly reversed profiles observed by SOHO/SUMER, Heinzel \emph{et
al}. (2001) also used multithread models and arrived at a remarkable
agreement with the observed line profiles and integrated intensities
for the first members of the Lyman series. Fontenla \emph{et al}.
(1996) considered a collection of threads in energy balance with the
surrounding corona. They also took into account ambipolar diffusion.
They found that ambipolar diffusion increases the emission in
Ly$\beta$ in comparison with other lines in the Lyman series leading
to a small Ly$\alpha$/Ly$\beta$ ratio compared to observations.
\\Engvold (1976), Engvold and Malville (1977) and Engvold, Malville, and Livingston (1978) observed
the fine structure of non-spot prominences with H$\alpha$
filtergrams. The size of the smallest prominence structures
increases with height above the chromosphere. Some prominences
contain structures close to 350 km, which is the spatial resolution
in these filtergrams. Some bright threads are visible for one hour
and longer. Their average line-of-sight velocity is about 30 km
s$^{-1}$ and their angular sizes are 1 Mm. Engvold (1978) observed
five hedgerow prominences with high spatial resolution and studied
some lines properties in them. He reported that the faint structures
appeared slightly hotter than the bright structures. Under good
seeing conditions, the quiescent prominences resolve into a fine
structure that consists of narrow threads and knots. These
structures are thought to arise from small-scale magnetic fields
embedded within the prominence, although a direct demonstration of
this connection has not yet been possible, for lack of observations
with sufficient spatial resolution (Zirker and Koutchmy, 1990). The
presence of fine structure must play an essential role in the
transfer of radiation, and possibly, heat conduction from the corona
(Zirker and Koutchmy, 1991). The cool threads may have their own
transition regions to the corona (see, \emph{e.g.}, Heinzel (2007)
or they may be embedded in a common transition region or they may be
isothermal threads each having different temperatures as suggested
by Poland and Tandberg-Hansen (1983). Pojoga, Nikoghossian, and
Mouradian (1998) studied the possible geometries of prominence fine
structures. Rovira \emph{et al}. (1994) and Fontenla \emph{et al}.
(1996) constructed thread modeling with ambipolar diffusion and they
assumed that each thread can be independently computed from its own
characteristics. The effect of radiative interaction of a number of
threads has been addressed by Heinzel (1989). Ajabshirizadeh and
Ebadi (2005) and Ajabshirizadeh, Nikoghossian, and Ebadi (2007) used
the method of addition of layers and calculated line profiles and
intensity fluctuations which mainly represent the fine structure
properties. Anzer and Heinzel (1999) presented slab models for
quiescent prominences in which both the condition of
magneto-hydrostatic equilibrium as well as Non-LTE are fulfilled.
They used the Kippenhahn-Schl$\ddot{u}$ter model (Kippenhahn and
Schl$\ddot{u}$ter, 1957) and deduced relations between the
components of the magnetic field, gas pressure, and prominence
width. Heinzel and Anzer (2001) constructed theoretical models for
vertical prominence threads which are in magneto-hydrostatic
equilibrium. Their models were fully two-dimensional and took the
form of vertically infinite threads hanging in a horizontal magnetic
field. They have shown that the Lyman profiles are more reversed
when seen along the lines, a behavior recently confirmed
observationally by Schmieder \emph{et al.} (2007). They showed the
effects of line-profile averaging over the fine structure threads
which are below the instrumental resolution. Heinzel and Anzer
(2003) and Heinzel, Anzer, and Gun\'{a}r (2005) deduced that
magnetically-confined structures in solar prominences exhibit a
large complexity in their shapes and physical conditions. Since
different Lyman lines and their line center, peak and wings are
formed at different depths within the thread, the Lyman series may
serve as a good diagnostic tool for thermodynamic conditions varying
from central cool parts to a prominence-corona transition region
(PCTR). They confirmed that the Lyman line profiles are more
reversed when seen across the field lines, compared to those seen
along the lines.
\\In Section 2 we present
the SUMER observations of Ly$\alpha$ and Ly$\beta$ and the full
information about the data that we used. Section 3 describes the data
processing methods and software that we used through this work.
Our results are presented in Section 4. It contains the cuts along the slit, Ly$\alpha$ and Ly$\beta$ profiles and their
ratios in different regions which correspond to the studied
prominence. Comparison of our results with previous works is done in Section 5 and conclusions are presented in Section 6.
\section{Observations}
SUMER is a high-resolution normal incidence spectrograph operating
in the range (780 -- 1610) {\AA}  (first order) and (390 -- 805)
{\AA} (second order). The spatial resolution along the slit is
1$^{''}$. The spectral resolution depends slightly on the
wavelength. It can vary from about 45 m{\AA} per pixel at 800 {\AA}
to about 41 m{\AA} per pixel at 1600 {\AA} (Wilhelm \emph{et al}.,
1995).
 A prominence situated at the limb in the north east quadrant was
observed with SUMER (detector A) on 25 May 2005. Figure 1 shows the
Ca {\sc II} K$_{3}$ (Meudon Observatory) and 304 {\AA} (SOHO/EIT)
images of the prominence at the time of our data. It shows that the
SUMER slit was well located for the prominence. The pointing
coordinates were X = 972$^{''}$, Y = 168$^{''}$. The slit which is
used was 0.3$^{''}$$\times$120$^{''}$. The observation was performed
from 16:23 UT to 16:26 UT for Ly$\alpha$ and from 16:20 UT to 16:22
UT for Ly$\beta$. It is clear that the Ly$\alpha$ and Ly$\beta$
observations are simultaneous within a couple of minutes.
 The exposure time for both of them was
 115 seconds. The spectral resolution is 43.67 m{\AA} and 44.37 m{\AA}
for Ly$\alpha$ and Ly$\beta$ respectively. The Ly$\alpha$
observations were performed in a special mode where the gain of the
detector was lowered in order to keep linearity for such an intense
signal.
\begin{figure}
\begin{center}
\centerline{ \psfig{file=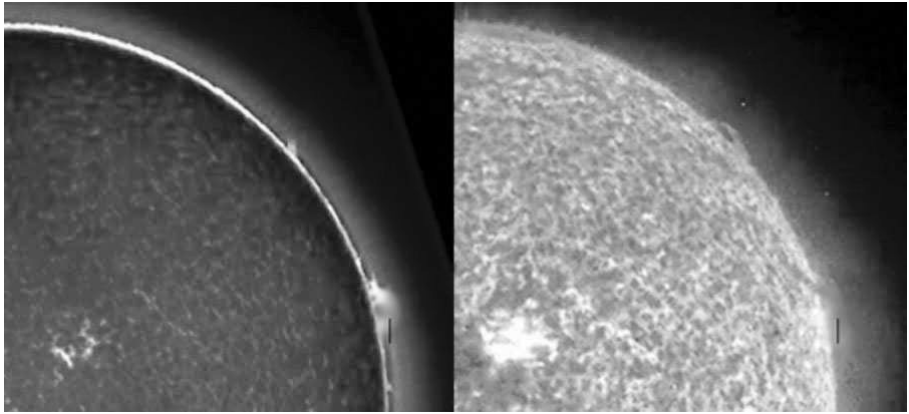,width=12cm,angle=0}}
\end{center}
\caption{Images of the studied prominence which were observed by
Meudon (Ca {$\sc$ II} K$_{3}$ observatory (left) and 304 {\AA}
SOHO/EIT (right) on 25 May 2005. The slit is in the North\,--\,South
direction.}
\end{figure}

\section{Data Processing}
The raw data have been initially processed applying the standard
procedures for geometric distortion, flat-fielding, and dead-time
correction which can be found in the Solar Software (SSW)
database. Once these corrections were applied, we performed the
radiometric calibration via the radiometry program in the SSW
environment. The specific intensity unit is W m$^{-2}$ sr$^{-1}$
\AA$^{-1}$ through this analysis. Because of the above-mentioned
special observing mode of the Ly$\alpha$ line, low signals may
well be underestimated by about ten per cent according to P.
Lemaire (private communication). This means that only the
Ly$\alpha$ wings may be affected.
\section{Results}
We calculated the integrated intensities
for both Ly$\alpha$ and Ly$\beta$ along the slit (Figure 2).
\begin{figure}
\begin{center}
\centerline{ \psfig{file=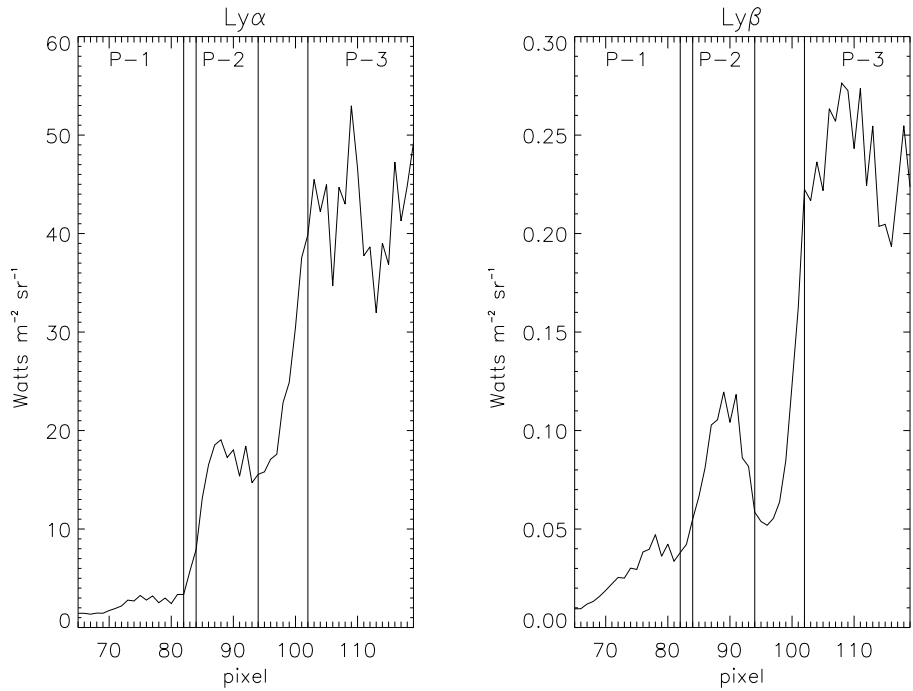,width=12cm,angle=0}}
\end{center}
\caption{The integrated intensity along the slit for Ly$\alpha$
(left) and for Ly$\beta$ (right). There exist
three regions with different intensities both in Ly$\alpha$ and
Ly$\beta$ lines.}
\end{figure}
Only pixels 60 to 120 are useable. The intensities were compared with previous
works and they are in the prominence range. We
present Ly$\alpha$ profiles in Figures 3 and 4 and Ly$\beta$
ones in Figures 5 and 6. The numbers above each profile
correspond to the position along the slit. Ly$\alpha$ line
profiles from pixels 60 to 84 are non-reversed but from pixels 85
to 119 they are reversed and the reversals are more apparent in
the last pixels. The intensities are increasing at the end of the
 slit in both lines. The reversed profiles in the case of
Ly$\beta$ are located from pixels 100 to 119. Some profiles are
red-shifted and some of them are blue-shifted in both lines. In some
cases a significant blue peak in Ly$\alpha$ coincides with a
significant red peak in Ly$\beta$ (\emph{e.g.} pixels 104 and 105),
but the opposite is true for pixels 110 and 115.
 \\As Figure 2 shows, there are three regions with
different intensities, so we decided to study the prominence in
three categories which are called P-1, P-2, and P-3.
\\We plotted Ly$\alpha$ and Ly$\beta$ average profiles separately for each region
in Figures 7 and 8. Ly$\alpha$ line is reversed in regions P-2 and P-3, but Ly$\beta$ is reversed only
in P-3. From P-1 to P-3 the average intensities and line widths increase in both lines.
\begin{figure}
\begin{center}
\centerline{ \psfig{file=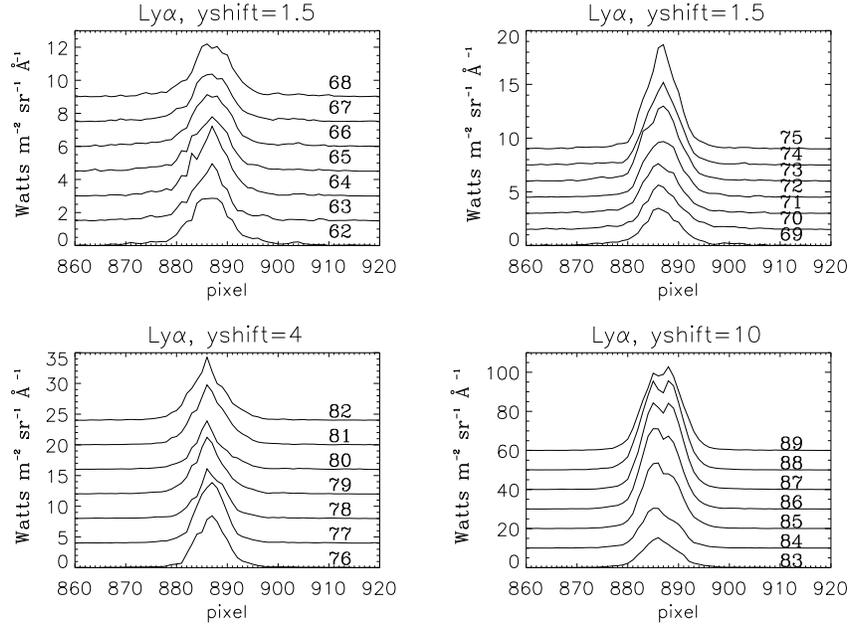,width=12cm,angle=0}}
\end{center}
\caption{Ly$\alpha$ profiles from pixels 62 to 89 along the slit.}
\end{figure}
\begin{figure}
\begin{center}
\centerline{ \psfig{file=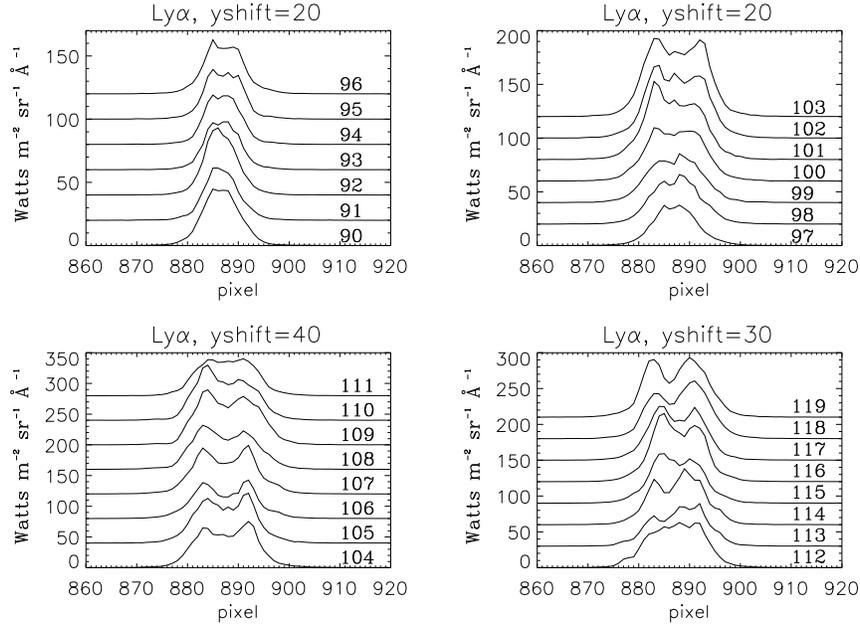,width=12cm,angle=0}}
\end{center}
\caption{Ly$\alpha$ profiles from pixels 90 to 119 along the
slit.}
\end{figure}
\begin{figure}
\begin{center}
\centerline{ \psfig{file=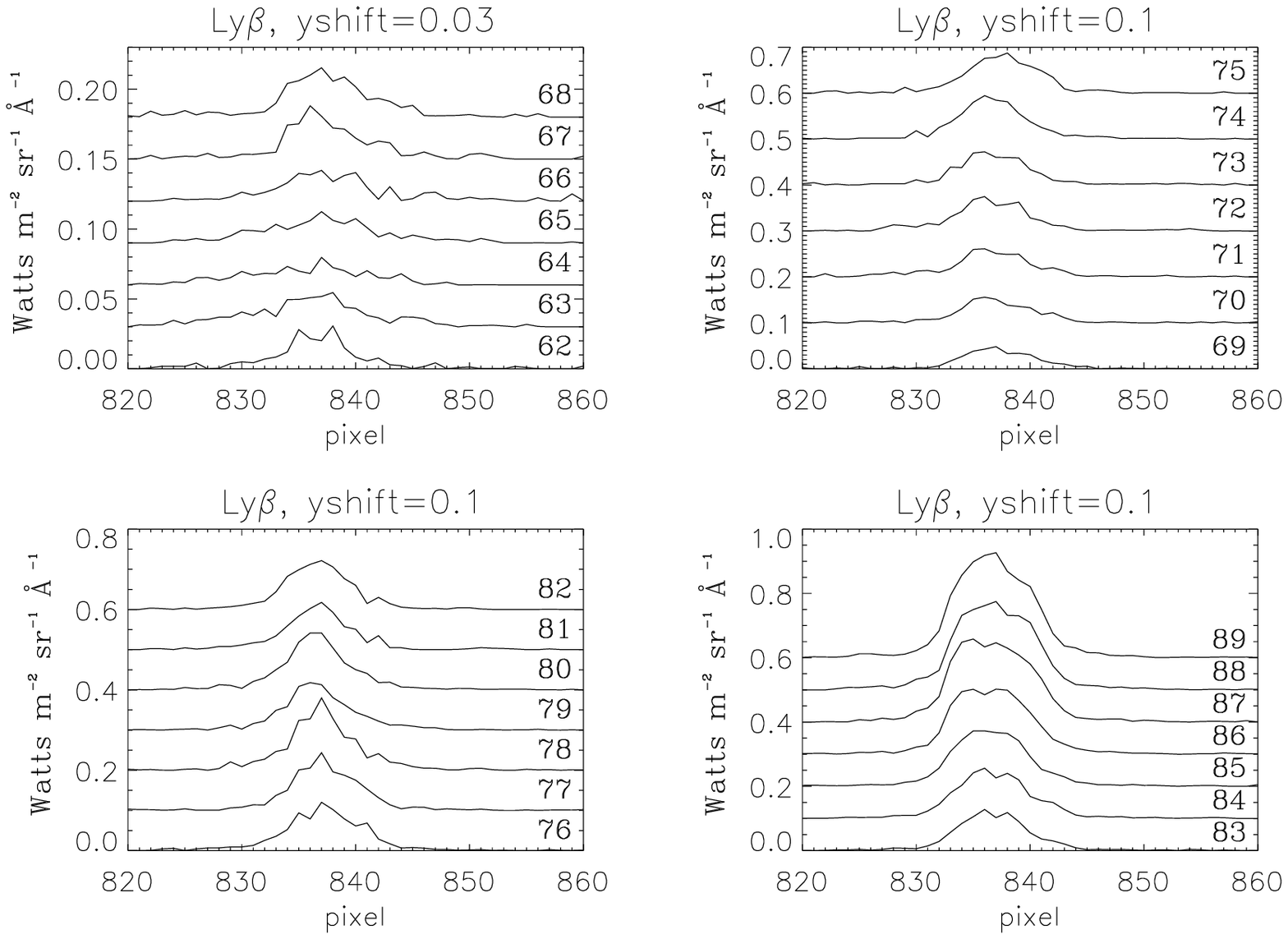,width=12cm,angle=0}}
\end{center}
\caption{Ly$\beta$ profiles from pixels 62 to 89 along the
slit.}
\end{figure}
\begin{figure}
\begin{center}
\centerline{ \psfig{file=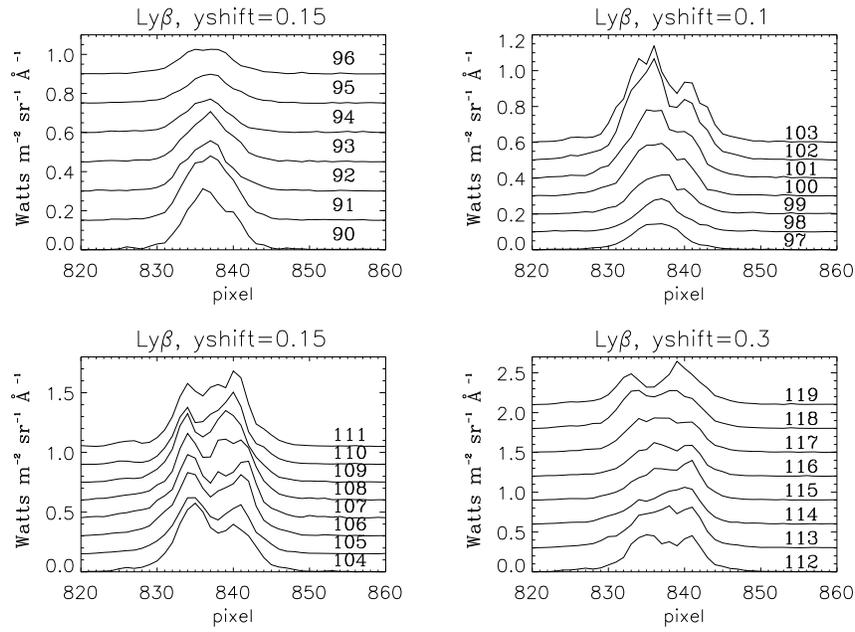,width=12cm,angle=0}}
\end{center}
\caption{Ly$\beta$ profiles from pixels 90 to 119 along the slit.}
\end{figure}
\begin{figure}
\begin{center}
\centerline{ \psfig{file=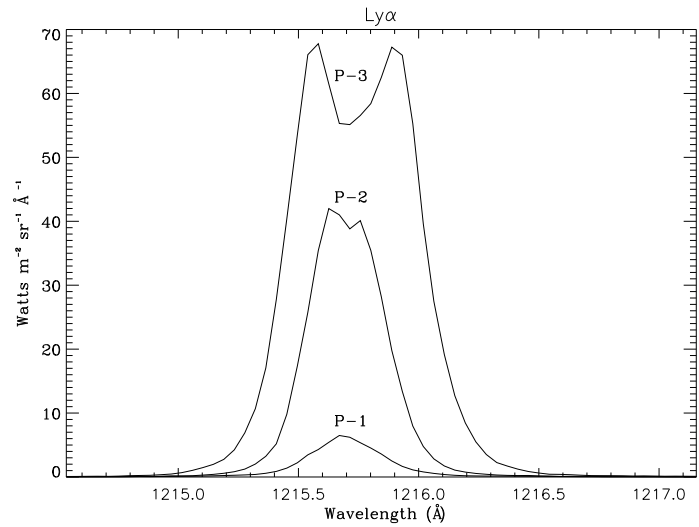,width=10cm,angle=0}}
\end{center}
\caption{Ly$\alpha$ average profiles for the three regions.}
\end{figure}
\begin{figure}
\begin{center}
\centerline{ \psfig{file=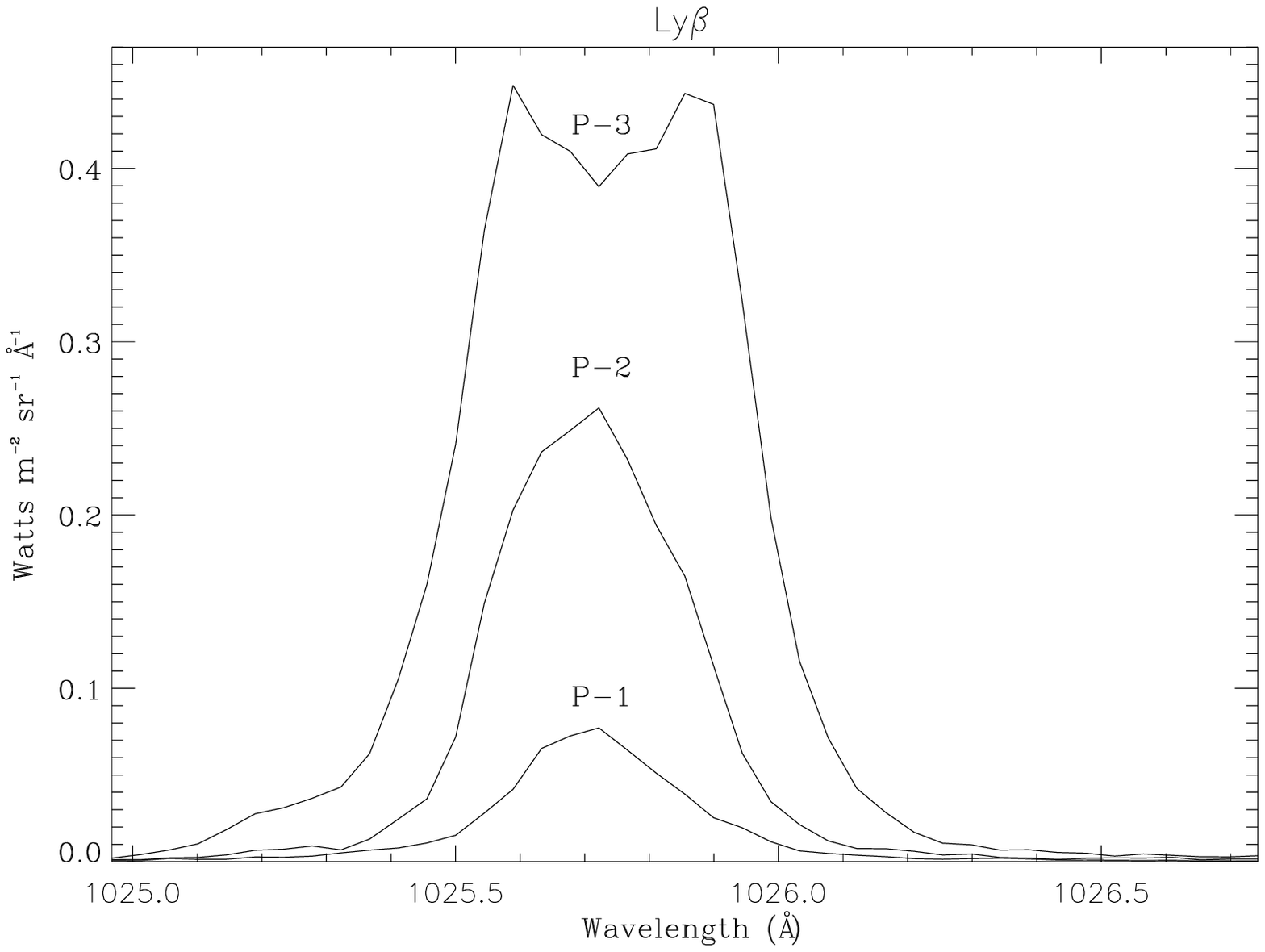,width=10cm,angle=0}}
\end{center}
\caption{Ly$\beta$ average profiles for the three regions.}
\end{figure}
\\The Ly$\alpha$/Ly$\beta$ ratio for the three
regions is presented in Figure 9.
The ratio in P-1 is smaller than in the other
regions. This ratio has fluctuations (12, 5, and 3 percent for P-1, P-2, and P-3 regions respectively)
 which are related to prominence fine structure. A summary of
our Ly$\alpha$ and Ly$\beta$ intensities and their ratio is
presented in Table I where they are compared to OSO 8
observations. Only the P-2 and P-3 Ly$\alpha$ and
Ly$\beta$ intensities are comparable (but not equal) to OSO 8
results. However, because of the high OSO 8 Ly$\beta$ intensity,
the OSO 8 Ly$\alpha$/Ly$\beta$ ratio is lower than ours by a factor three.
We are confident in our measurements made with high
statistics, specially in P-3. In terms of  photometric calibration, OSO 8 and SUMER
refer to the same disk absolute intensities (for Ly$\alpha$ and
Ly$\beta$ respectively). We can only speculate that the SUMER and OSO
8 observations were of very different structures.
\begin{table}[htbp]
\caption{The observed integrated intensities of Ly$\alpha$ and
Ly$\beta$ and their ratio.}
   \label{tabfreqoliver}
\vspace{0.6cm}
\begin{tabular}{c c c c}
   \hline
   Prominence region &  Ly$\alpha$ (W m$^{-2}$ sr$^{-1}$) & Ly$\beta$ (W m$^{-2}$ sr$^{-1}$) & Ly$\alpha$/Ly$\beta$  \\
   \hline
   P-1 &  \phantom{x}2.36 & 0.027 & \phantom{x}96  \\
   \hline
   P-2 &  15.87 & 0.089 & 183  \\
   \hline
   P-3 & 42.30 & 0.237 & 181 \\
     \hline
OSO 8 observations (Vial, 1982a) &  28\,--\,36 & 0.44\,--\,0.55 & \phantom{x}65  \\
   \hline
\end{tabular}
\end{table}
\begin{figure}
\begin{center}
\centerline{ \psfig{file=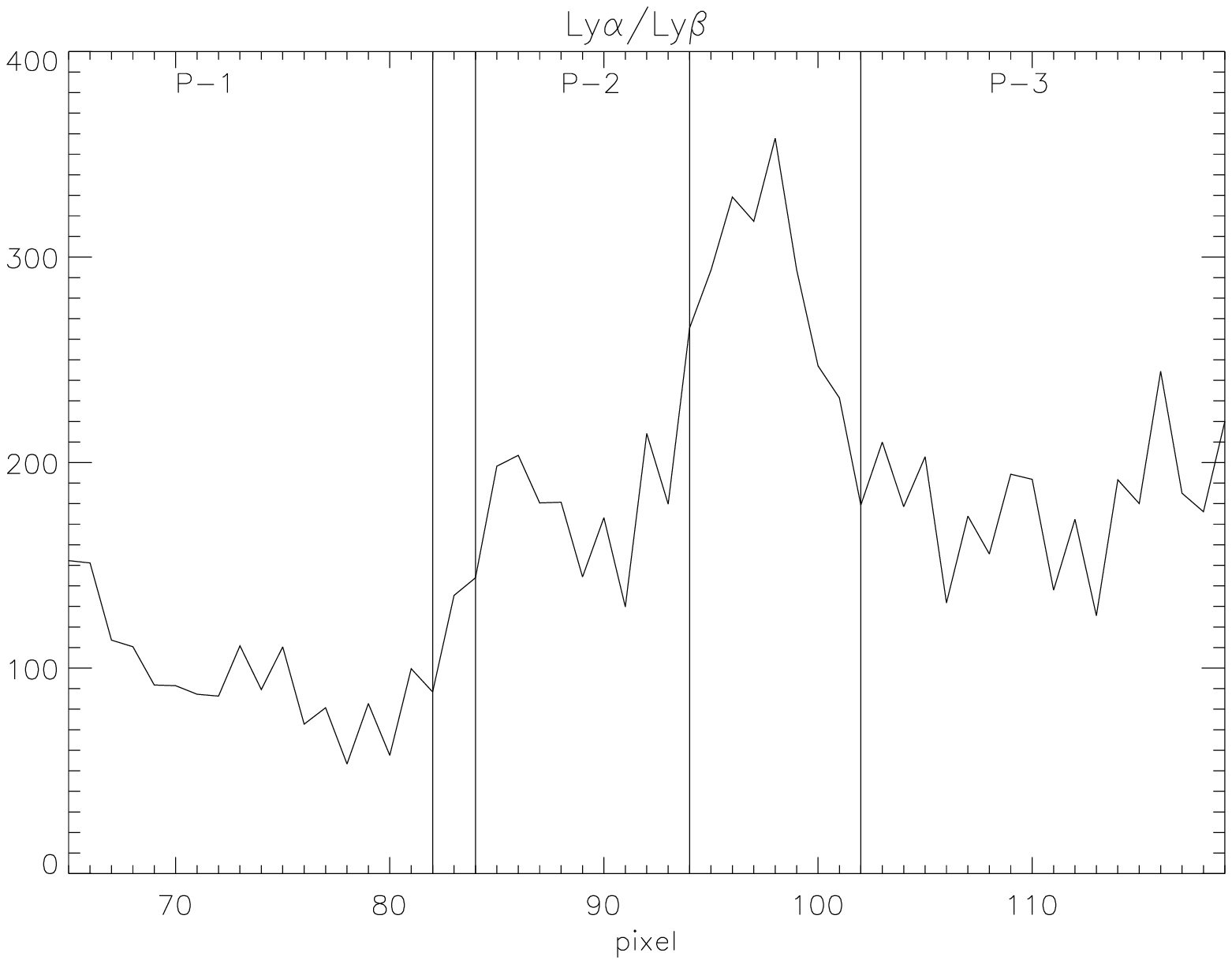,width=10cm,angle=0}}
\end{center}
\caption{Ly$\alpha$/Ly$\beta$ for the three regions (P-1, P-2 and P-3 from left to right).}
\end{figure}
\section{Comparison with the Fine Structure Modeling}
We observed different profiles with different intensities both in
Ly$\alpha$ and Ly$\beta$ which provide evidence of fine structuring
in solar prominences. We now compare these results (in terms of
integrated intensities), with non-LTE modeling. Non-LTE radiative
transfer modeling of prominences has been performed using
plane-parallel finite slabs by Gouttebroze, Heinzel, and Vial
(1993). We selected some pixels along the slit and managed to find
GHV models which match Ly$\alpha$ and Ly$\beta$ integrated
intensities. They are described in Table II where the first column
provides the pixel number and the last three columns describe the
relevant model. The range of fitting models is rather large, from
low temperature and low pressure models (8000 K and 0.01 dyn
cm$^{-2}$) to higher temperature and pressure ones (10000 K, 0.2 dyn
cm$^{-2}$), most of them being relatively thick (5000 km). We
noticed that bright regions are well fitted by models with low
temperature, high pressure and important geometrical thickness. But
these model agreements become rather illusory when one looks at the
actual shapes of the observed and computed profiles. The comparison
is performed in Table II where columns 2 and 3 each represent the
nature of the observed and modeled Ly$\alpha$ and Ly$\beta$
profiles, respectively. U means unreversed, R means reversed and F
flat-topped. One can see that the only agreement for both lines is
attained at pixels 107 and 110, where all profiles are actually
reversed. Let us also note that all other profiles of Table II are
NOT reversed, contrary to observations of Vial, 1982a. Such a
discrepancy could be interpreted in terms of the modeling of
prominence fine structures by Heinzel, Anzer, and Gun\'{a}r (2005):
These authors found that the Lyman line profiles are much more
reversed when seen across the field lines, compared to those seen
along the lines. The agreement between observed and modeled maximum
intensities (columns 4 and 5) is rather good. The agreement between
observed and modeled full widths at half maximum (FWHM) (columns 6
and 7) is generally excellent at the exception of a few ratios of
0.5. But maximum intensities and FWHM represent crude parameters of
lines profiles whose general observation/modeling agreement complies
with the integrated intensities agreement. Returning now to the
issue of integrated intensities, we also compared with systems of
thread models.
\\Fontenla and Rovira (1983, 1985) and Vial \emph{et al}. (1989)
developed non-LTE models of individual prominence threads including
a large number of narrow threads. They found that the computed
Ly$\alpha$ profiles are close to the observed ones, but the
Ly$\beta$ line intensities are too small and consequently, the
Ly$\alpha$/Ly$\beta$ ratio is high compared with that observed by
OSO 8 (Vial, 1982a). More recently, Fontenla \emph{et al}. (1996)
found that ambipolar diffusion increases the emission in Ly$\beta$
in comparison with other lines in Lyman series. However, the
ambipolar diffusion models give excessive Ly$\beta$ emission, viz.,
too small a Ly$\alpha$/Ly$\beta$ ratio compared with observations.
Table III presents a comparison of theoretical modeling results
including integrated intensities of Ly$\alpha$ and Ly$\beta$ and
their ratio with our results. Although we find some agreement with
the 1D models of Gouttebroze, Heinzel, and Vial (1993) and thread
modeling without ambipolar diffusion (Fontenla and Rovira, 1983,
1985; Vial \emph{et al}., 1989), we have no agreement with thread
modeling with ambipolar diffusion (Fontenla \emph{et al}., 1996).
\\The spatial variations of the integrated Ly$\alpha$ and Ly$\beta$ intensities and their
ratio could only be matched by a variety of 1D models, although the
detailed shapes of the profiles did not match (see Table II). The
variety of fitting models and their inadequacy for profile shapes
are a strong indication of the existence of a fine structure where
the cool core must be complemented with some kind of PCTR. Moreover,
the recent Swedish Solar Telescope observations of Lin \emph{et al}.
(2005) present
 unescapable evidence of fine structures as one can see by looking at the images. Actually,
the possibility of having an agreement with thread models without
ambipolar diffusion indicates that the temperature gradient at the
boundary of prominence threads may not be as strong as in the
chromosphere-corona transition region (and consequently the
ambipolar diffusion less efficient). Such a conclusion was already
reached by Chiuderi and Chiuderi-Drago (1991). Heinzel, Anzer, and
Gun\'{a}r (2005) also reached a similar conclusion in terms of
profile variations along (across) the magnetic field lines, where
conduction is parallel (perpendicular), respectively. Unfortunately,
because of insufficient geometrical information on the magnetic
structure of the prominence we observed and also because of the
projection effects when a prominence is seen at the limb, we cannot
assign our spatial observations to a particular orientation of our
line of sight with respect to the field lines.
\begin{table}[htbp]
\caption{Gouttebroze, Heinzel, and Vial (1993) 1Dimensional model
results which are in agreement with our observations. Column 1
presents the pixels along the SUMER slit and S1 and S2 are the
shapes of observed vs. modeled profiles for Ly$\alpha$ and Ly$\beta$
respectively. U means unreversed, F flat-topped and R reversed.
I$_{1}$ and I$_{2}$ are the ratio of observed maximum intensities to
the computed ones relevant to this model for Ly$\alpha$ and
Ly$\beta$ respectively. F1 and F2 are the ratio of the observed FWHM
to computed one for Ly$\alpha$ and Ly$\beta$ lines respectively.
R$_{\mbox{OC}}$ is the ratio of the observed Ly$\alpha$/Ly$\beta$
integrated intensities to the computed ones. Three last columns
present the model parameters (temperature (K), pressure (dyn
cm$^{-2}$) and geometrical thickness (km)).}
   \label{tabfreqoliver}
\vspace{0.6cm}
\begin{tabular}{c c c c c c c c c c c c}
   \hline
    \textrm{Pixel} & \textrm{S}1 & \textrm{S}2 & I$_{1}$ & I$_{2}$ & F1 &
   F2 & R$_{\mbox{OC}}$ & T & P & $\Delta z$ \\
   \hline
85 & U/R & F/U & 1.0 & 0.8 & 1.1 & 1.3 & 0.8 & \phantom{x}8000 & 0.01 & 5000 \\
 \hline
86 & U/R & U/U & 1.1 & 0.9 & 0.8 & 1.4 & 1.0 & \phantom{x}8000 & 0.01 & 1000 \\
  \hline
90 & U/R & U/R & 1.0 & 1.0 & 0.5 & 1.1 & 0.5 & \phantom{x}8000 & 0.20 & 1000  \\
 \hline
93 & U/R & U/U & 1.0 & 1.1 & 0.5 & 1.0 & 0.6 & \phantom{x}8000 & 0.01 & 5000 \\
 \hline
99 & F/R & U/F & 1.1 & 0.9 & 1.1 & 1.3 & 0.9 & \phantom{x}8000 & 0.20 & \phantom{x}200 \\
   \hline
100 & F/R & U/R & 1.0 & 0.9 & 1.5 & 1.3 & 1.4 & 15000 & 0.01 & 5000  \\
   \hline
 102 & F/R & U/F & 1.0 & 0.8 & 0.8 & 1.3 & 0.6 & \phantom{x}8000 & 0.20 & \phantom{x}200  \\
     \hline
 107 & R/R & R/R & 1.0 & 1.0 & 1.2 & 1.0 & 0.8 & 10000 & 0.05 & 5000  \\
     \hline
 110 & R/R & R/R & 0.9 & 0.9 & 1.5 & 1.5 & 0.8 & 10000 & 0.20 & \phantom{x}200  \\
     \hline
111 & F/R & F/R & 1.1 & 0.8 & 0.6 & 1.1 & 0.7 & \phantom{x}8000 & 0.20 & 5000  \\
     \hline
\end{tabular}
\end{table}
\begin{table}[htbp]
\caption{Integrated intensities of Ly$\alpha$ and Ly$\beta$ and
their ratio (1: Our results, 2: Gouttebroze, Heinzel, and Vial
(1993), 3: Fontenla and Rovira (1983, 1985) and Vial \emph{et al}.
(1989) 4: Fontenla \emph{et al}. (1996)).}
   \label{tabfreqoliver}
\vspace{0.6cm}
\begin{tabular}{c c c c}
   \hline
Reference &  Ly$\alpha$ (W m$^{-2}$ sr$^{-1}$) & Ly$\beta$ (W m$^{-2}$ sr$^{-1}$) & Ly$\alpha$/Ly$\beta$  \\
   \hline
   SUMER observations (1) & \phantom{xx}2.36\,--\,42.3 &  0.027\,--\,0.237 & 96\,--\,180  \\
   \hline
   One-dimensional Modeling (2) &  \phantom{x}7.50\,--\,45 &  0.037\,--\,0.29 & 90\,--\,400  \\
   \hline
  Thread modeling without A.D. (3) & 13.60\,--\,55 & 0.1\,--\,1.3 & 42\,--\,111 \\
     \hline
   Thread modeling with A.D. (4) & \phantom{x}10.80\,--\,38.6 & 0.59\,--\,7.36 & 2.3\,--\,18.5 \\
     \hline
\end{tabular}
\end{table}
\section{Conclusions}
We have presented nearly simultaneous Ly$\alpha$ and Ly$\beta$
profiles obtained in a prominence with the SUMER spectrograph on
SOHO. Significant variability of these profiles on scales as small
as \textbf{1$''$} is present. Reversed and unreversed profiles were
obtained in both lines with behaviors which differ from one line to
the other (\emph{e.g.}
 significant blue peak in Ly$\alpha$ coinciding with a significant red
peak in Ly$\beta$). Although the number of observed profiles is
limited to about 60, making a proper statistical analysis
impossible, we believe that such spectral signatures result from
fine structuring of prominences. In bright regions of the
prominence, the Ly$\alpha$ intensity is larger than the OSO-8 value,
and the Ly$\beta$ intensity is lower. We have some agreements with
one-dimensional models and thread models without ambipolar
diffusion, but thread modeling with ambipolar diffusion gives high
Ly$\beta$ intensity and as a result too small a ratio. Such a result
indicates that the temperature gradient at the boundary of threads
may not be as strong as in the 1D PCTR models or that the magnetic
field direction has a profound influence on the profiles (as shown
by Heinzel, Anzer, and Gun\'{a}r (2005)). Actually, a detailed
comparison between some observed and 1D-modeled profiles(see Table
II) supports this idea. Note that the same data as used in this
paper were recently analyzed in terms of 2D fine structure models in
magnetohydrostatic equilibrium and fitting the line profiles of the
Lyman series has pointed to multithreads seen across the magnetic
field lines giving rise to reversed profiles (Gun\'{a}r \emph{et
al}. 2007). A combination of spectral and imaging information in the
Lyman series (in particular the Ly$\alpha$ and Ly$\beta$ lines)
would be a still more efficient tool for deriving the actual fine
structure of prominences.
\\\\\textbf{Acknowledgments:} The authors thank
P. Lemaire, F. Baudin, P. Boumier, K. Wilhelm, and K. Bocchialini
for useful comments on the data processing, and J. Leibacher for his
help in improving the manuscript. They warmly thank the anonymous
referee for help in improving the document. The observations took
place in the frame of the 15th MEDOC Campaign. The authors thank all
MEDOC Campaign participants, especially K. Wilhelm, B. Schmieder,
and P. Schwartz for their contribution to the Campaign. SUMER is
financially supported by DLR, CNES, NASA, and the ESA PRODEX program
(Swiss contribution). SOHO is a mission of international cooperation
between ESA and NASA.

\end{article}

\begin{thebibliography}{}

\bibitem{}  Ajabshirizadeh, A. and Ebadi, H.: 2005, \textit{J. Quant. Spectrosc. Radiat. Transfer }\textbf{95}, 127.

\bibitem{}  Ajabshirizadeh, A., Nikoghossian, A.G., and Ebadi, H.: 2007, \textit{J. Quant. Spectrosc. Radiat. Transfer  }
\textbf{103}, 351.

\bibitem{}  Anzer, U. and Heinzel, P.: 1999, \emph{Astron.
Astrophys.} \textbf{349}, 794.

\bibitem{}  Chiuderi, C. and Chiuderi-Drago F.: 1991, \textit{Solar Phys.}\textbf{132}
, 81.

\bibitem{}  Engvold, O.: 1976, \textit{Solar Phys. }\textbf{49}, 283.

\bibitem{}  Engvold, O.: 1978, \textit{Solar Phys. }\textbf{56}, 87.

\bibitem{}  Engvold, O. and Malville, J.M.: 1977, \textit{Solar Phys. }\textbf{52}, 369.

\bibitem{}  Engvold, O., Malville, J.M., and Livingston, W.: 1978, \textit{Solar Phys. }\textbf{60}, 57.

\bibitem{}  Fontenla, J. M.  and Rovira, M.: 1983, \textit{Solar Phys. }\textbf{85}, 141.

\bibitem{}  Fontenla, J. M.  and Rovira, M.: 1985, \textit{Solar Phys. }\textbf{96}, 53.

\bibitem{}  Fontenla, J.M.,  Rovira, M.,  Vial, J.-C., and Gouttebroze, P.: 1996,
\textit{Astrophy. J. }\textbf{466}, 496.

\bibitem{}  Gouttebroze, P.,  Heinzel, P., and Vial, J. C.: 1993,
\textit{Astron. Astrophys. }\textbf{99}, 513 (GHV).

\bibitem{}  Gun\'{a}r, S., Heinzel, P., Schmieder, B., and Anzer, U.: 2007, In: Heinzel, P., Dorotovic, I., Rutten, R.J.
(eds.), \emph{The Physics of Chromospheric Plasmas, \emph{Astron.
Soc. Pacific Conf. Ser., San Francisco}} \textbf{368}, 317.

\bibitem{}  Heinzel, P.: 1989, \textit{Hvar Obs. Bull.} \textbf{13}, 317.

\bibitem{}  Heinzel, P.: 2007, In: Heinzel, P., Dorotovic, I., Rutten, R.J. (eds.), \emph{The Physics of Chromospheric
Plasmas, \emph{Astron. Soc. Pacific Conf. Ser., San Francisco}}
\textbf{368}, 271.

\bibitem{}  Heinzel, P. and Anzer, U.: 2001, \emph{Astron. Astrophys.}
\textbf{375}, 1082.

\bibitem{}  Heinzel, P. and Anzer, U.: 2003, In: Hubeny, D., Mihalas, D., Werner, K.
(eds), \emph{Stellar Atmosphere Modeling, \emph{Astron. Soc. Pacific
Conf. Ser., San Francisco}} \textbf{288}, 441.

\bibitem{}  Heinzel, P., Anzer, U., and Gun\'{a}r, S.:
2005, \textit{Astron. Astrophys. } \textbf{442}, 331.

\bibitem{}  Heinzel, P., Schmieder, B., Vial, J.-C., and Kotrc, P.: 2001,
\textit{Astron. Astrophys. }\textbf{370}, 281.

\bibitem{}  Kippenhahn, R. and Schl$\ddot{u}$ter, A.: 1957,
\textit{Z. Astrophys. }\textbf{43}, 36.


\bibitem{}  Lin, Y., Engvold, O., Rouppe, L., Wiik, J.-E., and Berger, T.-E.: 2005,
\textit{Solar Phys. }\textbf{226}, 239.

\bibitem{}  Pojoga, S., Nikoghossian, A. G., and Mouradian, Z.:
1998, \textit{Astron. Astrophys. }\textbf{332}, 325.

\bibitem{}  Poland, A.I. and Tandberg-Hansen, E.: 1983, \textit{Solar Phys. }\textbf{84}, 63.

\bibitem{}  Rovira, M. G., Fontenla, J. M., Vial, J.-C., and Gouttebroze, P.:
1994, In: Rusin, V., Heinzel, P., and Vial J.-C. (eds.), \emph{Solar
Coronal Structures, \emph{IAU Colloq., VEDA Publish. Comp.,
Bratislava}} \textbf{144}, 315.

\bibitem{}  Schmieder, B., Heinzel, P., Vial, J.-C., and Rudway,
P.: 1999, \textit{Solar Phys. }\textbf{189}, 109.

\bibitem{}  Schmieder, B., Gun\'{a}r, S., Heinzel, P., and Anzer, U.: 2007, \textit{Solar Phys. }\textbf{241}, 53.

\bibitem{}  Vial, J.-C.: 1982a, \textit{Astrophy. J. }\textbf{253}, 330.

\bibitem{}  Vial, J.-C.: 1982b, \textit{Astrophy. J. }\textbf{254}, 780.

\bibitem{}  Vial, J.-C., Rovira, M.G., Fontenla, J.M., and Gouttebroze, P.: 1989,
\textit{Hvar Obs. Bull., }\textbf{13(1)}, 331.

\bibitem{}  Wilhelm, K., Curdt, W., Marsch, E., Sch$\ddot{u}$hle, U.,
Lemaire, P., Gabriel, A., \emph{et al}.: 1995, \textit{Solar Phys.}
\textbf{162}, 189.

\bibitem{}  Zirker, J.B. and Koutchmy, S.: 1990, \textit{Solar Phys.}\textbf{127}, 109.

\bibitem{}  Zirker, J.B. and Koutchmy, S.: 1991, \textit{Solar Phys.}\textbf{131}, 107.

\end{thebibliography}
\end{document}